\documentclass[epj]{svjour} 
\usepackage{latexsym}
\usepackage[reqno,tbtags]{amsmath}
\usepackage{epsfig}
\usepackage{multirow}
\begin{document}

\newcommand{\wn}{$\mathrm{cm}^{-1}$ }
\newcommand{\wnf}{$\mathrm{cm}^{-1}$}
\newcommand{\vp}{v\'{}}
\newcommand{\vpA}{v$_{\mathrm A}\ $}
\newcommand{\vpc}{v$_{\mathrm c}\ $}
\newcommand{\vs}{v$_{\mathrm X}$}
\newcommand{\Js}{J\H{}}
\newcommand{\Xs}{X$^1\mr\Sigma_{\mathrm g}^{\mathrm +}$ }
\newcommand{\Bs}{B$^1\mr\Sigma_{\mathrm u}^{+}$ }
\newcommand{\As}{A$^1\mr\Sigma_{\mathrm u}^{+}$ }
\newcommand{\as}{a$^3\mr\Sigma_{\mathrm u}^{+}$ }
\newcommand{\sigu}{^1\rm\Sigma_{\mathrm u} }
\newcommand{\Piu}{^3\rm\Pi_{\mathrm u} }
\newcommand{\tsigu}{^3\rm\Sigma_{\mathrm u} }
\newcommand{\cs}{c$^3\rm\Pi_{\mathrm u}$ }
\newcommand{\rcm}{\mbox{cm$^{-1}$}}
\newcommand{\JJ}{\frac{\hbar^2}{2\mathrm{\mu R^2}}X}
\newcommand{\JJO}{\frac{\hbar^2}{2\mathrm{\mu R^2}}(X+1)}
\newcommand{\Jop}{\hat{\mathrm{J}}}
\newcommand{\Lop}{\hat{\mathrm{L}}}
\newcommand{\Sop}{\hat{\mathrm{S}}}
\newcommand{\mR}{\mathrm{R}}
\newcommand{\mr}{\mathrm}
\newcommand{\vnew}{v$_\mathrm{new}$}

\title{Study of coupled states for the (4s$^2$)$^1$S + (4s4p)$^3$P asymptote of Ca$_2$.}
\author{O. Allard \inst{1} \and St. Falke \inst{1} \and A. Pashov\inst{2} \and O. Dulieu  \inst{3} \and H. Kn\"ockel \inst{1} \and E. Tiemann \inst{1}
}                     

\offprints{E. Tiemann, tiemann@iqo.uni-hannover.de}     

\institute{Institut f\"ur Quantenoptik, Universit\"at Hannover, Welfengarten 1, 30167 Hannover, Germany \and Department of Physics, Sofia University, 5 James Bourchier blvd, 1164 Sofia, Bulgaria \and Laboratoire Aim\'e Cotton, CNRS, B\^at 505, Campus d'Orsay, 91405, Orsay Cedex, France}

\date{Received: date / Revised version: date}

\abstract{ The coupled states \As ($^1$D + $^1$S), \cs ($^3$P + $^1$S) and \as ($^3$P + $^1$S) of the calcium dimer are investigated in a laser induced fluorescence experiment combined with high-resolution Fourier-transform spectroscopy. A global deperturbation analysis of the observed levels, considering a model, which is complete within the subspace of relevant neighboring states, is performed using the Fourier Grid Hamiltonian method. We determine the potential energy curve of the \As and \cs states and the strengths of the couplings between them. The \cs and \as states are of particular importance for the description of collisional processes between calcium atoms in the ground state $^1$S$_0$ and  excited state $^3$P$_1$ applied in studies for establishing an optical frequency standard with Ca.} 


\maketitle

\section{Introduction}

\begin{figure}[ht]
\centering
\resizebox{0.95\columnwidth}{!}{
  \includegraphics{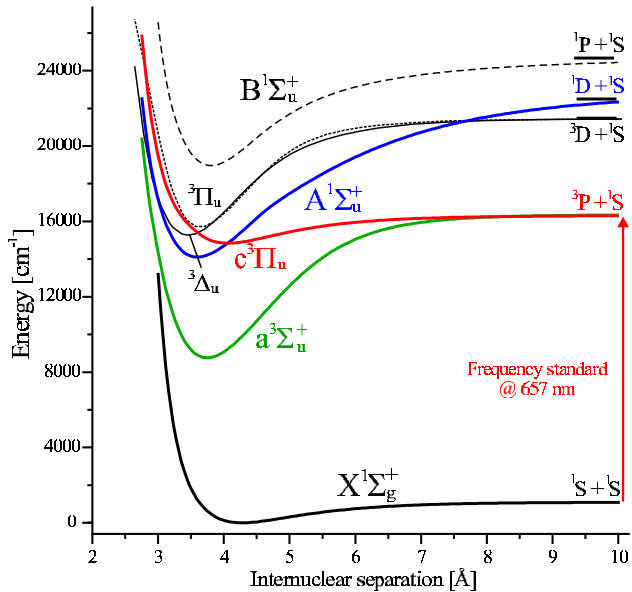}}
\caption{Simplified potential scheme of the calcium dimer. The potential curves of the \As and \cs states are determined in this study. The curve of the \Xs state was published in \cite{Allard2} and the curve of the \Bs state is given in \cite{ThesisAllard}. The potential curve of the \as, $^3\mr\Delta_{\rm u}$, $^3\mr\Pi_{\rm u}$ states have been obtained by Czuchaj\textit{ et al.} \cite{Czuchaj} by \textit{ab initio} method.}
\label{potentials}    
\end{figure}

It was recently demonstrated that an optical frequency standard using an ensemble of ultra- cold calcium atoms, which is probed on the intercombination transition $^3$P$_1$ (m$_{\mr j}$~=~0) $\leftarrow$~$^1$S$_0$ by a high precision laser, has the potential to exceed the microwave cesium clock in stability and accuracy \cite{CaPTB,CaNIST}. The performance of such an optical clock depends on the possibility to reduce or to correct for the effects of any disturbances during the laser interrogation of the transition. Collisions between calcium atoms lead to a frequency shift of the transition, which limits the achievable accuracy. Therefore, the investigation of the collisional processes between atoms in the involved $^1$S$_0$ and  $^3$P$_1$ states is of particular importance for the improvement of the accuracy of the frequency standard. Such high accuracy is required for testing fundamental theories for the quest of combining general relativity with quantum mechanics or to search for cosmological variation of natural constants.\\
In this perspective we have reported previously on the study of the interactions between two $^1$S$_0$ calcium atoms \cite{Allard1,Allard2}. In this article we present our investigations on the interactions between one atom in the ground state and one excited atom in the $^3$P$_{1,2}$ state. With the knowledge of the molecular potentials correlated to the $^3$P$_1$ + $^1$S$_0$ dissociation limit, we would be able to calculate photoassociation spectra near the intercombination line and might gain a better understanding of trap losses occurring during the cooling cycle on this transition. \\ 
Like for the atomic intercombination line, the observation of the \cs and \as states by a direct excitation from the ground state is rather inefficient. However, the \cs state is strongly coupled via spin-orbit interaction to the \As state, which is reachable from the ground state. The \as is indirectly coupled to the \As state by spin-orbit and rotational coupling via the \cs state. Therefore, we can investigate the A state and observe its perturbed level structure in order to obtain information about all potentials of the coupled states and the coupling strengths. Generally, we are observing also levels of the triplet states taking advantage of their singlet character.\\
Therefore, we aim to determine the potential energy curves of the coupled-states system \As ($^1$D + $^1$S), \cs ($^3$P + $^1$S) and \as ($^3$P + $^1$S) (see figure \ref{potentials}). We will perform a global deperturbation analysis of the observed levels, which enables us to treat all the perturbations of the levels, simultaneously. This method will provide physical parameters for the potentials and the coupling strengths while local deperturbation methods will only give phenomenological parameters \cite{Lisdat}. We collect spectroscopic information of a whole set of rovibrational levels of the coupled states. The shape of the potential curves will be adjusted by a fitting procedure.\\ 
This article is organized as follows. In section \ref{PE}, we review the previous investigations of the coupled states. The experimental set-up and the spectroscopic observations realized by laser induced fluorescence and filtered laser excitation spectroscopy are reported in section \ref{ESSO}. We derive the Hamiltonian of the coupled states in section \ref{Ham}, and describe, in section \ref{GlobDep}, the global deperturbation analysis for determining the potential energy curves of the coupled states and coupling interactions. In section \ref{Disc}, we discuss the results of the deperturbation analysis and conclude in section \ref{Conc} by a comparison with earlier results and by proposing further studies.

\section{Previous experiments of the coupled system} \label{PE}
The \As $-$ \Xs system was observed by Bondybey and English \cite{Bond84} in a supersonic jet created from the vaporization of calcium metal by a pulsed laser. The spectroscopic interrogation was realized by a pulsed dye laser system limiting the experimental uncertainty
  to a few tenths of a wave number. 
Several bands starting from low rovibrational levels of the ground state were observed. \\
Hofmann and Harris published the results of a more systematic investigation of the \As - \Xs system \cite{Hof84,Hof86}.
They performed laser induced fluorescence spectroscopy in a heat pipe oven, and applied the filtered laser excitation technique (FLE) to observed band heads. 
They stated an experimental uncertainty of 0.01 \wnf. From the observation of P-R doublets they assigned the progression to the \As $-$ \Xs system, and measured 720 lines corresponding to 340 different rovibrational levels of the coupled states A and c. 
Their observed and assigned transitions are available through the original publication \cite{Hof86}, therefore we used their data in our own analysis. 
Their assignment of the lowest observed vibrational level of the A state is v$_{\mathrm {H\& H}}$ = 7.  This high lying number 7 leaves room for different assignments. Therefore, vibrational assignment will be re-established using the additional and more precise spectroscopic data we have measured. This notation v$_{\mathrm {H\& H}}$ for the vibrational quantum number v will be used here on to distinguish the earlier from a new assignment given below.

\section{Experimental set-up and spectral observations} \label{ESSO}

\begin{figure}
\centering
\resizebox{1.0\columnwidth}{!}{
  \includegraphics{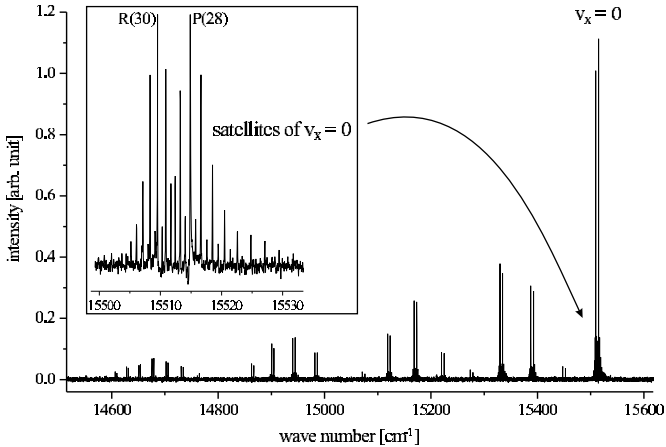}}
\caption{Vibrational progression obtained from the excited (v$_{\mathrm {H\&H}}$ = 12, 29)$ \leftarrow $(0, 28) transition of \As $-$ \Xs.}
\label{v11sat}    
\end{figure}

The calcium dimer was formed in a stainless steel heat pipe oven, characteristics of which are described in a previous publication \cite{Allard1}. The pipe was filled with approximately 5g of $^{40}$Ca ($99,5\%$ purity) and with argon as buffer gas at a pressure around 50 mbar. We operated the oven at temperatures between $1240$ K and 1275 K.  
The sample was irradiated along the pipe axis by different laser sources. A frequency stabilized, linear dye laser (Coherent 599)  pumped by an argon ion laser (Innova 400), using Pyridine 1 and DCM dyes, was run in single mode with a typical output power of 70 mW. Transitions of the \As - \Xs system in the intervals from 13850 \wn to 14550 \wn and from 14900 \wn to 15650 \wn are excited. We collected the induced fluorescence through a Fourier-transform spectrometer (Bruker IFS 120HR) by a broad band photomultiplier. This total collection results to the LIF data set. 
Each recorded spectrum exhibits a single progression of rotational doublets as expected from the selection rule $\mr\Delta \mr J = \pm 1$ for the rotational quantum number J for $\mr\Sigma \leftrightarrow \mr\Sigma$ transitions. Few spectra show several progressions due to the overlap of exciting transitions with the laser frequency within their Doppler linewidth. Because the rotational spacing of the doublets is determined by the ground state and thus well known \cite{Allard2}, the rotational assignment of the observed lines is obtained without ambiguity. The strongest lines are surrounded by collisionally induced satellites. These satellite lines permit to measure the position of neighboring rovibrational levels around the excited levels. Figure \ref{v11sat} shows the P-R doublets progression from the excitation transition (v$_{\mathrm {H\&H}}$ = 12, J = 29)$ \leftarrow $(v''=0, J''=28) at 15514.843 \wnf. On the insert a zoom around the fluorescence lines P(28) and R(30) of \vs = 0 shows the relatively large number of collision induced satellites (24 satellites corresponding to 12 different J values). 

\begin{figure}
\centering
\resizebox{1.0\columnwidth}{!}{
  \includegraphics{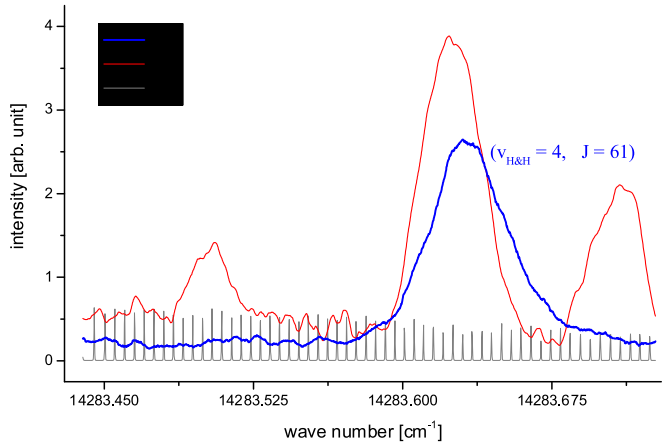}}
\caption{(4, 61) $\leftarrow$ (11, 62) Ca$_2$ calcium line recorded with I$_2$ absorption lines (reversed here) and peaks of a marker cavity FPI. The traces of Ca$_2$ and I$_2$ have been smoothed. The fluorescence detection window was set to the transition (4, 61) $\rightarrow$ (3, 62) at 14284.9 \wnf and a width of $\delta\nu = 2$\wnf.} 
\label{v4J61}
\end{figure}

\begin{figure}[h!]
\centering
\resizebox{1.0\columnwidth}{!}{
  \includegraphics{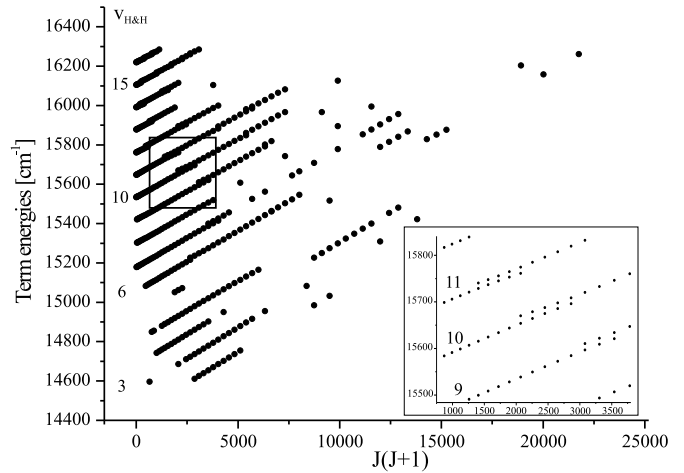}}
\caption{Term energies with respect to the minimum of the ground state potential \cite{Allard2} obtained in this study and in the study of Hofmann and Harris. A zoom of the region delimited by a square is presented in the insert. Extra levels are clearly visible for several J values, for instance for the rotational ladder of v$_{\mathrm {H\& H}}$ = 9, 10 and 11 shown in the insert.} 
\label{term}
\end{figure}

Due to the possible excitation of different classes of velocities depending on the frequency of the laser, a Doppler shift can occur. Before recording with the Fourier spectrometer, we ensured that the excitation was tuned to the maximum of emission for the selected A-X transitions for which we directed the fluorescence, emitted in the forward direction of the laser, to a 1~m monochromator (GCA/McPherson Instruments). It was used with a band pass of about 2 to 5~\wn width. The center of the frequency window was set to the position of a line of the induced fluorescence progression with favorable Franck-Condon factors, involving typically a low lying vibrational level of the ground state. The light passing through the monochromator was detected using a broad band Hamamatsu photomultiplier (R928). The detection of the fluorescence intensity allowed setting and fixing the laser frequency to the maximum of emission. \\
In this way, fluorescence progressions have been observed from the systematic excitation of rotational levels of v$_{\mathrm {H\&H}}$~= 1, 2, 3, 6, 7, 8, 9, and 11. Additional rotational levels of v$_{\mathrm {H\&H}}$~= 5, 10, 12, and 13 have been observed. The levels below v$_{\mathrm {H\&H}}$ = 7 are levels, which were not observed by Hofmann and Harris \cite{Hof86}.\\
Special care should be put on the experimental uncertainties. By setting the laser to the center of the line, we reduced the possible Doppler shifts to a magnitude of about $0.006$~\wn $\approx$ $180$ MHz. The drift of the frequency of the stabilized laser was less than 10 MHz per hour, which is sufficiently smaller than the desired setting of the center frequency given above and certainly also smaller than the Doppler width of the lines in this frequency region ($\sim 1.3$ GHz $\approx$ $0.043$ \wnf), to allow a stable excitation during the time of recording ($\sim 20$ min corresponding to 20 scans). 
The resolution of the Fourier-transform interferometer was chosen to be 0.05 \wnf. With a triangular apodization, the instrument line width is 0.05 \wnf. Since in single mode operation of the laser, only a selected velocity class is excited, no Doppler broadening is expected, and velocity changing collisions play no role under these conditions. The lifetime of the A state is 57$\pm$5 ns \cite{Bond84} giving a homogeneous broadening of about 6 $\times$10$^{-4}$ \wnf, which is negligible compared to the instrumental broadening.
The broadening due to the size of the aperture of the instrument ($\sim$ 1.3 mm) is in the order of 0.025 \wnf.
Taking the considered effects together, we expect a line width of 0.056 \wnf. The measured widths of the lines with good signal-to-noise ratio (SNR), i.e. higher than 10, were $\sim 0.062$ \wnf. Collisional broadening due to the high temperature and the buffer gas is probably responsible for the slightly larger value. We have evaluated the relative frequency uncertainties of the observed lines to be 10 times smaller than their line widths. For lines with lower SNR, we estimated a higher uncertainty proportional to the reduced SNR. \\
In the frequency range accessible with our dye lasers the excitation and detection of fluorescence progressions from v$_{\mathrm {H\&H}}$ =~4 have less favorable Franck-Condon factors (4 times smaller than for v$_{\mathrm {H\&H}}$~= 3). In consequence the number of collisional satellites with good signal-to-noise ratio was reduced to very few. In this case, the Fourier- transform spectroscopy does not help to observe a wide portion of the vibrational band by the satellites.  \\
We applied the filtered laser excitation spectroscopy (FLE) by recording the emission of Ca$_2$ through the monochromator. The temperature of the oven was chosen to 1240 K since the enhancement of the collisional transfer of population is not needed for the FLE spectroscopy. For absolute and relative frequency calibration, the absorption lines of iodine from a 60 cm long cell (heated up to 870 K), and marker cavity peaks, with 149.7 MHz spectral spacing, were recorded. A precise description of the technique is given in \cite{Allard2}. 
Rotational levels from J = 27 to 77 were observed for v$_{\mathrm{H\& H}}$ = 4 in this way.
The (4, 61) $\leftarrow$ (11, 62) measured Ca$_2$ line is presented in the figure \ref{v4J61} as example.
We used the IodineSpec software program \cite{Iod} to calibrate the spectra. It provides a prediction of I$_2$ transitions with accuracy better than 25 MHz in this spectral region. The width of the calcium lines was 0.042(1)~\wn corresponding to the Doppler width of 0.04~\wn and a residual broadening of about 0.012(3)~\wnf. The width of the iodine lines was in the order of 0.04~\wn depending on the unresolved hyperfine structure. The absolute experimental uncertainty is determined by the precision of the estimation of the  I$_2$ and Ca$_2$ line centers. The signal-to-noise ratio was sufficiently high for the I$_2$ and Ca$_2$ lines to allow a determination of their centers better than 0.004~\wn for each of them. The final accuracy of the line position was then estimated to be 0.006 \wnf. \\
We have constructed term energies from our observed transitions and those reported by Hofmann and Harris \cite{Hof86} in adding the term energies of the involved levels of the well-known ground state potential \cite{Allard2}. From the LIF data set, the term energy of a level of the A state is obtained several times since the observed spectrum consists of a rovibrational progression from that selected upper level. Within each observed progression we have averaged the constructed term energies weighted by their relative uncertainties. To the resulting uncertainty from this average, we added the absolute uncertainty 0.009 \wnf, which is the absolute uncertainty of the frequency given by the Fourier-transform spectrometer, to obtain the experimental uncertainty of one term energy. Then the term energies of the same levels from different recordings have been averaged, weighted with their respective experimental uncertainty. Finally, the total uncertainty varies from 0.009 \wn to 0.0380 \wnf. The data of Hofmann and Harris consist of P and R transitions for 340 levels of the A state. We  averaged the two values obtained for each upper level. \\
We mention here that the levels observed by the filtered laser excitation technique have an experimental uncertainty smaller than those obtained by LIF and those of Hofmann and Harris' data set. Only level v$_{\mathrm {H\&H}}$~= 4 has been observed with this technique. In order to avoid a too high weight for the fits on this single vibrational level compared to all the others, we have increased their uncertainties to 0.015~\wnf which is equal to the typical uncertainty of the LIF line with good signal-to-noise ratio. \\
Figure \ref{term} shows the obtained term energies with respect to the minimum of the ground state potential \cite{Allard2}. The progressions as function of J(J+1) for each vibrational levels of the A state are clearly visible. In the insert, the term energies of v$_{\mathrm{H\& H}}$ = 9, 10, and 11 around J = 45, i.e J(J+1) = 2070, are presented as examples, to show the perturbed rotational structure of the levels of the \As state. Around a crossing with a vibrational level of the \cs state, the mixing between the two states becomes important. The levels of the triplet c state get sufficient singlet character to become observable yielding extra lines in the spectrum. The energy positions of such levels are important since they carry a lot of information about the c state and the coupling strength.  \\
We have compared the data set from Hofmann and Harris to our own measurements by calculating the differences between term energies formed with their and our observed transitions. The overlapping set consists of 64 common levels. The data set from Hofmann and Harris is shifted in average by 0.037 \wn to higher energy compared to our measured set. The standard deviation of this shift is $0.038$ \wnf. This spread is larger than our experimental accuracy and the one claimed by Hofmann and Harris (0.01 \wnf). We assign to all their data an error of $0.038$ \wnf. Later on, the global treatment shows that the data set of Hofmann and Harris should be lowered by $0.028$ \wnf\ to give the best fit. We remove from the data set of Hofmann and Harris the 64 common levels and keep ours, since they have been observed with a higher accuracy. The total data set consists of 502 term energies.\\ 
The LIF and FLE spectroscopic techniques enabled us to observe six lower vibrational levels of the A state compared to the data set observed by Hofmann and Harris. Local perturbations have been identified at lower energy than the energy of the lowest perturbation observed in their study, which they proposed to be caused by \vpc = 0 of \cs(0$^+$). Therefore, these observations show already the necessity of a revision of their vibrational assignment for the \cs state. Since more information on the lower vibrational levels of the \As has been collected, the vibrational assignment of this state should be examined, too. 
A list of all known transition frequencies of the system (\As, c$^3\mr{\Pi_u}$) $-$ \Xs and the derived term energies are available in the supplementary \emph{ Online material}.

\subsection{Vibrational and rotational assignments}
From the first inspection of the data we assumed that no local perturbation was found for v~$_{\mathrm{H\& H}}$ $<$ 4, which was later confirmed by the complete analysis. The coupling between the A state and the c state affects the vibrational levels below v$_{\mathrm{H\& H}}$ = 4 only by a global shift of the rovibrational levels towards lower energies (see \cite{Lisdat}). The variation of the shifts with v$_{\mathrm{H\& H}}$ and J is monotonic compared to the local perturbation at higher v$_{\mathrm{H\& H}}$-values. These shifts are also much smaller than the vibrational spacing of the levels. We can thus consider that although a one-channel fit of such levels will not provide a good reproduction of energy level positions, it will nevertheless be sufficient for a good estimate of Franck-Condon factors (FCF). We tested different assignments of the A state by comparing the intensity pattern of the fluorescence progressions to the FCF calculated from the fitted potential corresponding to each assignment. 
\begin{figure}
\centering
\resizebox{1.0\columnwidth}{!}{
  \includegraphics{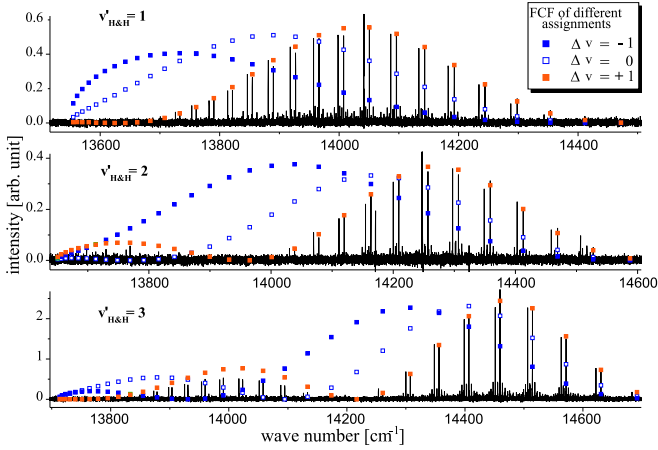}}
\caption{Calculated FCF for different assignment of the A state levels, compared to the observed fluorescence progressions obtained by the excitation of the transitions (v$_{\mathrm{H\& H}}$ = 1, 61) $\leftarrow$ (8, 62), (v$_{\mathrm{H\& H}}$ = 2, 57) $\leftarrow$ (6, 58) and (v$_{\mathrm{H\& H}}$ = 3, 47) $\leftarrow$ (4, 46). Agreement is obtained for $\mr\Delta$v$ = +1$.} 
\label{fcf-assign}
\end{figure} 
We tested the following assignments $\mr\Delta$v = $-1$, $\mr\Delta$v = 0, $\mr\Delta$v = $+1$ and $\mr\Delta$v = $+2$ with $\mr\Delta$v = \vnew $-$ v$_{\mathrm{H\& H}}$. The fluorescence progressions obtained by the excitation of the transitions (v$_{\mathrm{H\& H}}$ = 1, 61) $\leftarrow$ (8, 62), (v$_{\mathrm{H\& H}}$ = 2, 57) $\leftarrow$ (6, 58) and (v$_{\mathrm{H\& H}}$ = 3, 47) $\leftarrow$ (4, 46) are compared to the predicted Franck-Condon factors for the different assignments, and normalized to the most intense line of each progression. These comparisons are presented in figure \ref{fcf-assign}. In this figure the assignment corresponding to the $\mr\Delta$v = $+2$ has been omitted for saving space. Convincing agreement is obtained for $\mr\Delta$v = $+1$. Therefore all observed vibrational levels have been shifted by +1 for the A state. This correct assignment will be used in the remainder of this article as \vpA = v$_{\mathrm{H\& H}}$ + 1. \\
To determine the vibrational assignment of the c state, we have to find the position of the lowest local perturbation on the rotational progressions of the vibrational levels of the A state. This perturbation is then assigned to a crossing with \vpc = 0 of the \cs ($\mr\Omega = 0$) state. In figure \ref{Bvcst}, an estimate of the effective rotational constant $B_v$ = $\mr\Delta$(E)/(4J$-$2) is presented for levels \vpA = 2 to 8. $\mr\Delta$(E) is the spacing between two observed consecutive rotational levels (J $-$ 2) and J. We see on the lower graph that the $B_v$-values of \vpA = 7 and 8 abruptly decrease due to local perturbations. Despite that only two $B_v$-values are obtained for \vpA = 6, the significantly lower values clearly indicate the presence of a local perturbation.  
 \begin{figure}
\centering
\resizebox{1.0\columnwidth}{!}{
  \includegraphics{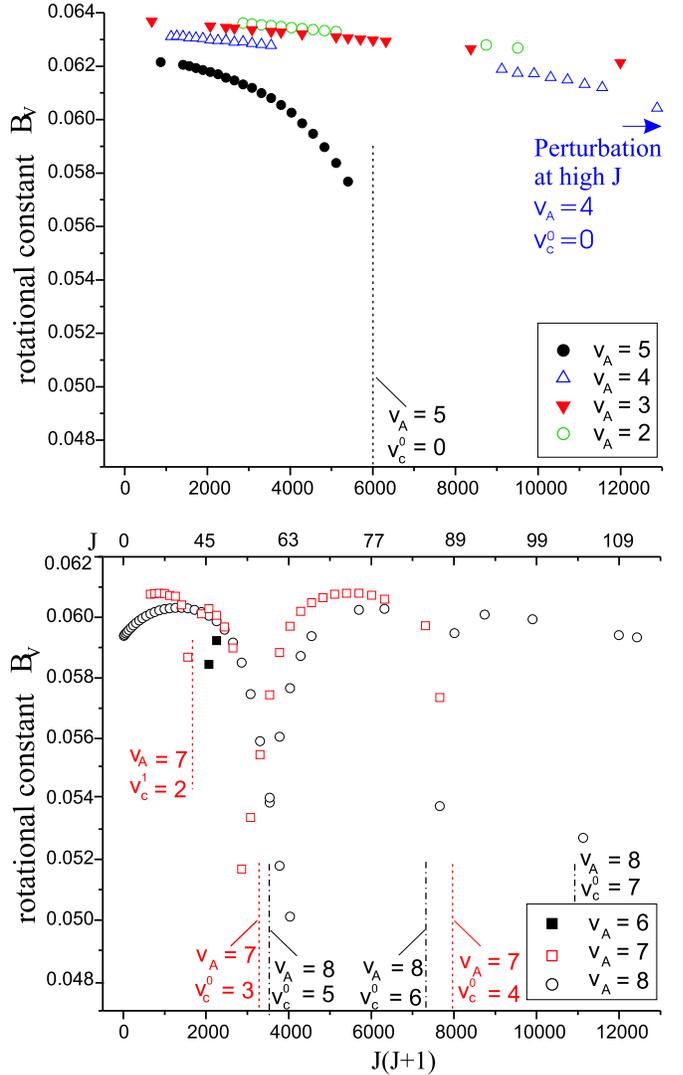}}
\caption{Effective rotational constants for low vibrational levels. The vertical dashed lines indicate positions where the rotational ladder of a vibrational level \vpA of the A state crosses the rotational ladder of a vibrational level v$_{\mr c}^\mr\Omega$ of one $\mr\Omega$-component of c$^3\mr\Pi_{\mr u}$.} 
\label{Bvcst}
\end{figure}
If we compare the magnitude of the variation of the rotational constants for \vpA = 3 and 2 with the higher vibrational levels, effects of local perturbations on these two levels should be visible in the range of observed rotational numbers, despite the observation gaps in the series of levels. But, such a local perturbation is not present. Instead we have smooth and slight decrease of the effective rotational constant. A relatively fast change of the effective rotational constant is visible for \vpA = 5 at J values approaching 77. A moderate decrease of the $B_v$-values for \vpA = 4 is observed at high J indicating the onset of a local perturbation for this vibrational level. Considering that no local perturbations are present for \vpA = 3 and 2 for J $<$ 109 and assuming that the same vibrational level of the c state causes the two local perturbations on \vpA = 4 and 5, we conclude that these two local perturbations are the lowest in energy, and are caused by the level v$_{\mr c}^{\mr\Omega =0}$ = 0 of the \cs state, which is the lowest in the whole manifold of $^3\mr\Pi_{\mr u}$. These observations and assumptions will lead to a consistent picture of the perturbations as we will show in section \ref{GlobDep}. \\ 
Even if there are no crossings with the c state levels of the lower levels of the A state they are globally shifted to lower energy due to their couplings to all the levels above. Consequently, a global treatment of the coupled states is required. \\
Figure \ref{Adata} gives the range of vibrational and rotational quantum numbers observed in this work and those reported by Hofmann and Harris. 
 \begin{figure}
\centering
\resizebox{1.0\columnwidth}{!}{
  \includegraphics{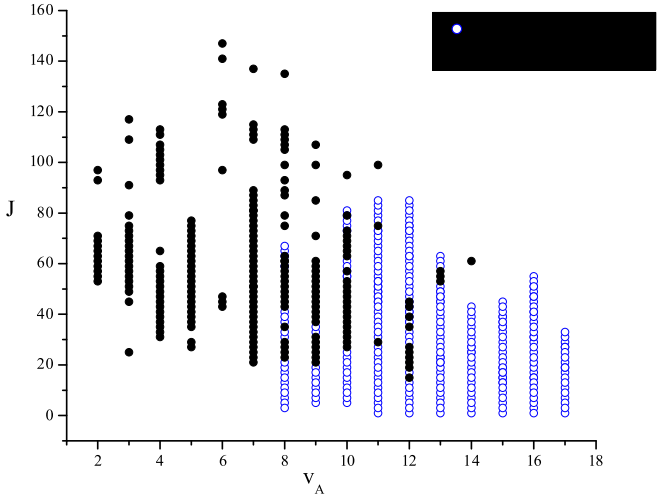}}
\caption{Range of quantum numbers of observed levels following the revised assignment. The black dots are from this work. The blue open circles are the data from Hofmann and Harris \cite{Hof86}. Levels corresponding to extra lines have received the same vibrational quantum numbers as those of the A state levels.} 
\label{Adata}
\end{figure}
\\

\section{The Hamiltonian for the coupled A, c and a states} \label{Ham}

The Hamiltonian of a diatomic molecule in the body-fixed frame $xyz$ can be expressed as:
\begin{equation}\label{Hamilt}
{\mbox{H}}={\mbox{V}}^{\mathrm{BO}} +{\mbox{T}}^{\mathrm{N}} + {\mbox{H}}^{\mathrm{ROT}} + {\mbox{H}}^{\mathrm{REL}},
\end{equation} 
where
${\mr{V}}^{\mathrm{BO}}$ is the Born-Oppenheimer potential matrix, and ${\mr{T}}^{\mathrm{N}}$ is the nuclear radial kinetic energy operator. ${\mbox{H}}^{\mathrm{ROT}}$ is the rotational Hamiltonian given by: 
{\small \begin{equation}
 \begin{split}
 {\mr{H}}^{\mathrm{ROT}}&= \frac{1}{2\mu\mR^2}[\hat{\mR}]^2\\&
 = \frac{1}{2\mu\mR^2}[(\Jop^2 - \Jop^2_z) + (\Lop^2 - \Lop^2_z) + (\Sop^2 - \Sop^2_z) +\\&
 \underbrace{(\Lop_+\Sop_- + \Lop_-\Sop_+)}_{\mathrm{spin-electronic}} - \underbrace{(\Jop_+\Lop_- + \Jop_-\Lop_+)}_{\mathrm{L-uncoupling}} - \underbrace{(\Jop_+\Sop_- + \Jop_-\Sop_+)}_{\mathrm{S-uncoupling}}]    
 \end{split}
\end{equation}}
where $\Jop = \hat{\mR} + \Sop + \Lop$ is the total angular momentum (the nuclear spin of calcium I =0), $\hat{\mR}$ is the nuclear rotation operator, $\Sop$ is the total electronic spin, and $\Lop$ is the total electronic orbital angular momentum.
We use the ladder operators $\hat{\mathrm{O}}_{\pm} =\hat{\mathrm{O}}_x \pm i\,\hat{\mathrm{O}}_y$ where $\hat{\mathrm{O}}$
stands for $\Jop$, $\Lop$ and $\Sop$. H$^{\mathrm{REL}}$ contains the spin-orbit interaction H$^{\mathrm{SO}}$, the spin-spin interaction H$^{\mathrm{SS}}$ and the spin-rotation interaction ${\mathrm{H}}^{\mathrm{SR}}$.\\
Here we limit ourselves to the coupled manifold of states \As correlated to the (4s4s$^1$S + 4s3d$^1$D) asymptote, \cs and \as both correlated to the (4s4s$^1$S + 4s4p$^3$P) asymptote, see figure \ref{potentials}. Other states dissociating at the asymptote 4s4s$^1$S + 4s3d$^3$D can couple to the considered states (see fig. \ref{potentials}) but they are expected to lie above the triplet manifold according to the \textit{ab initio} calculations of \cite{Czuchaj}. Their coupling will be much less pronounced due to the absence of resonance effects, and will not be included in our analysis. Consequently we will derive potentials that might contain these small influences, and, in this respect, should be considered as effective potentials. Two potentials, the $^3\mr\Pi_u$ ($^1$S + $^3$D ) and the $^3\mr\Delta_u$ ($^1$S + $^3$D ) are deep enough according to \textit{ab initio} calculations to overlap and cross the repulsive branch of the c state (see fig. \ref{potentials}). But this overlap appears at the very bottom of these potentials, where the density of levels is small. The magnitude of the perturbations will be most likely small, because of low Franck-Condon factors between these states and the states A and c. The influence of these states is neglected in our deperturbation analysis. For a similar reason the influence of the \as state is small since it is much deeper than the c state to which it is coupled. Near the asymptotic limit ($^3$P + $^1$S) where the energy spacing between the potential of the c state and the a state becomes smaller than the spin-orbit interaction the coupling between these states is not negligible. The way to include this state in our analysis will be discussed in section \ref{4s}. 

\subsection{Basis functions}

In Hund's case (a) the angular momentum basis functions are $|\alpha,\mr J,\rm{S},\mr\Omega,\mr\Lambda,\mr\Sigma>$, where $\mr\Lambda$ and $\mr\Sigma$ are respectively the projections onto the molecular axis of the total electronic orbital angular momentum and of the total electronic spin momentum,  $\mr\Omega = \mr\Lambda + \mr\Sigma$, and  $\alpha$ stands for all other state labels (e.g. electronic configuration, g/u symmetry).
The two indistinguishable calcium atoms are bosons with nuclear spin zero, therefore levels of the ground state exist only for even J and have ($e$) symmetry, where the $(^e_f)$ basis functions are defined by the inversion symmetry ($\pm$)($-1$)$^\mr{J}$ for molecules with even number of electrons \cite{Brown}. Thus, only rotational levels with ($e$) symmetry and odd J of the \As, \cs and \as system can be observed via the excitation from the ground state \Xs. In addition, perturbation can only occur between levels of same $(^e_f)$ symmetry. Therefore, in the following we only consider the subspace of levels with ($e$) symmetry. The properly symmetrized functions with ($e$) symmetry and odd values of J are the following (\cite{Brion}):
\begin{eqnarray}
&| ^{2\mr{S}+1}\mr\Lambda_\mr\Omega,\mr J,e> &=\nonumber \\
&(2)^{-1/2}[|\alpha,\mr{J},\mr\Lambda,\mr{S},&\mr\Sigma,\mr\Omega>-(-1)^{-\mr{S}}|\alpha,\mr{J},\mbox{-}\mr\Lambda,\mr{S},\mbox{-}\mr\Sigma,\mbox{-}\mr\Omega>]  \nonumber\\
\end{eqnarray}
Then, the normalized wave functions for the five involved states and J odd are:
{\small 
\begin{eqnarray}
|A\sigu^+,\mr{J},e> & = & |\alpha,J,0,0,0,0> \nonumber \\ 
|c\mr\Piu (0^+),\mr{J},e> & = & (2)^{-1/2}[|\alpha,\mr{J},1,1,\mbox{-1},0>+|\alpha,\mr{J},\mbox{-1},1,1,0>] \nonumber  \\
|c\mr\Piu (1),\mr{J},e> & = & (2)^{-1/2}[|\alpha,\mr{J},1,1,0,1>+|\alpha,\mr{J},\mbox{-1},1,0,\mbox{-1}>] \nonumber  \\
|c\mr\Piu (2),\mr{J},e> & = & (2)^{-1/2}[|\alpha,\mr{J},1,1,1,2>+|\alpha,\mr{J},\mbox{-1},1,\mbox{-1},\mbox{-2}>]  \nonumber \\
|a\tsigu^+(1),\mr{J},e> & = & (2)^{-1/2}[|\alpha,\mr{J},0,1,1,1>+|\alpha,\mr{J},0,\mbox{1},\mbox{-1},\mbox{-1}>]  \nonumber\\ \label{wfunc}
\end{eqnarray}}
For simplicity the labels ($e$) and $\alpha$ will not be repeated hereafter.

\subsection{Matrix elements}

Matrix elements for the different terms of the Hamiltonian of eq.(\ref{Hamilt}) can be found in \cite{Brion} and an example of application in the case of K$_2$ in \cite{Lisdat}. 
We give here the matrix elements concerning our case first for the diagonal elements and second for the off-diagonal elements.\\
Diagonal matrix elements of $\hat{\mr{H}}^{\mathrm{ROT}}$ are:
\begin{eqnarray}
<\mr{J},\mr\Lambda,\mr{S},\mr\Sigma,\mr\Omega|{\mbox{H}}^{\mathrm{ROT}}|\mr{J},\mr\Lambda,\mr{S},\mr\Sigma,\mr\Omega> =\hspace{1.3cm}&  \nonumber \\
 \frac{\hbar^2}{2\mu\mR^2} \times [\mr{J}(\mr{J}+1) - \mr\Omega^2 + \mr{S}(\mr{S}+1) - \mr\Sigma^2]&
\end{eqnarray}
Diagonal matrix elements for the different contributions in H$^\mr{REL}$ are the following.\\
For the spin-orbit interaction ${\mathrm{H}}^{\mathrm{SO}}$: 
\begin{equation}
<\mr{J},\mr\Lambda,\mr{S},\mr\Sigma,\mr\Omega|{\mathrm{H}}^{\mathrm{SO}}|\mr{J},\mr\Lambda,\mr{S},\mr\Sigma,\mr\Omega> = \mr\Lambda\mr\Sigma A(\mR)
\end{equation}
A(R) is the spin-orbit function to be determined. The rotational interactions couple states within the same multiplicity. \\ 
For the other terms of the relativistic Hamiltonian ${\mr{H}}^{\mr{REL}}$ which contribute with a much smaller magnitude than the previous one we have for the spin-rotation Hamiltonian \cite{Brion}:\\
\begin{equation}
{\mr{H}}^{\mathrm{SR}} = \gamma(\mR)\ {\hat{\mR}}\cdot{\Sop}= \gamma(\mR)\ (\Jop - \Lop - \Sop)\cdot{\mr{\hat S}}
\end{equation}
 \begin{eqnarray}\label{gamma0}
<\mr{J},\mr\Lambda,\mr{S},\mr\Sigma,\mr\Omega| &{\mr{H}}^{\mathrm{SR}}|\mr{J},\mr\Lambda,\mr{S},\mr\Sigma,\mr\Omega> =&  \nonumber \\
 &\gamma(\mR)\ [\mr\Sigma^2 - \mr{S}(\mr{S}+1)]&
\end{eqnarray} 
And for the spin-spin Hamiltonian ${\mbox{H}}^{\mathrm{SS}}$:
\begin{equation}\label{SS}
{\mbox{H}}^{\mathrm{SS}} = \epsilon(\mR)\ (3\Sop^2_z - \Sop^2)
\end{equation}  
 leading to:
\begin{eqnarray}\label{SS1}
<\mr{J},\mr\Lambda,\mr{S},\mr\Sigma,\mr\Omega|&{\mr{H}}^{\mathrm{SS}}|\mr{J},\mr\Lambda,\mr{S},\mr\Sigma,\mr\Omega> = & \nonumber \\
&\epsilon(\mR)\ (3\Sigma^2 - \mr{S}(\mr{S}+1))&
\end{eqnarray}
$\gamma$(R) and $\epsilon$(R) are unknown quantities in our problem.\\
For the off-diagonal matrix elements of ${\mbox{H}}^{\mathrm{ROT}}$, we distinguish the different contributions of the Hamiltonian. 
The S-uncoupling operator is:
\begin{equation}
{\mr{H}}^{\mathrm{JS}} = -\frac{1}{2\mu\mR^2}(\Jop_+\Sop_- + \Jop_-\Sop_+)\ ,
\end{equation}
it contributes as:
$$<\mr{J},\mr\Lambda,\mr{S},\mr\Sigma,\mr\Omega|{\mr{H}}^{\mathrm{JS}}|\mr{J},\mr\Lambda,\mr{S},\mr\Sigma-1,\mr\Omega+1> =\hspace{3.5cm}$$
\begin{equation}
-\frac{\hbar^2}{2\mu\mR^2}\times \sqrt{\mr{J}(\mr{J}+1) - \mr\Omega(\mr\Omega+1)}\times \sqrt{\mbox{S}(\mbox{S}+1) - \mr\Sigma(\mr\Sigma-1)}
\end{equation}
These contributions are symbolized by $\mr\Gamma_{JS}^{0}$ for $\mr\Omega = 0$ and by $\mr\Gamma_{JS}^{1}$ for $\mr\Omega = 1$.\\ 
The spin-electronic term is:
\begin{equation}
{\mr{H}}^{\mathrm{LS}} = \frac{1}{2\mu\mR^2}(\Lop_+\Sop_- + \Lop_-\Sop_+)\ ,
\end{equation}
it gives the contribution:
\begin{equation}
 \begin{split}
<\mr{J},&\mr\Lambda,\mr{S},\mr\Sigma,\mr\Omega|{\mr{H}}^{\mathrm{LS}}|\mr{J},\mr\Lambda+1,\mr{S},\mr\Sigma-1,\mr\Omega> =\\
&\frac{\hbar^2}{2\mu\mR^2}\times \sqrt{\mr{S}(\mr{S}+1) - \mr\Sigma(\mr\Sigma-1)}\times L(\mR)
 \end{split}
\end{equation}
and is symbolized by $\mr\Gamma_{LS}$ for $\mr\Omega = 1$ and the function $L$(R) is the expectation value of the $\Lop_\pm$ operator:
\begin{eqnarray} \label{L}
L(\mR)&=<\mr{J},\mr\Lambda+1,\mr{S},\mr\Sigma,\mr\Omega+1|\Lop_+|\mr{J},\mr\Lambda,\mr{S},\mr\Sigma,\mr\Omega>\nonumber \\
&=<\mr{J},\mr\Lambda,\mr{S},\mr\Sigma,\mr\Omega|\Lop_-|\mr{J},\mr\Lambda+1,\mr{S},\mr\Sigma,\mr\Omega+1>\,\,
\end{eqnarray}
The L-uncoupling operator is:
\begin{equation}
{\mr{H}}^{\mathrm{JL}} = -\frac{1}{2\mu\mR^2}(\Jop_+\Lop_- + \Jop_-\Lop_+)
\end{equation}
and gives contributions:
\begin{eqnarray}
<&\mr{J},\mr\Lambda,\mr{S},\mr\Sigma,\mr\Omega|{\mr{H}}^{\mathrm{JL}}|\mr{J},\mr\Lambda-1,\mr{S},\mr\Sigma,\mr\Omega+1> =&\nonumber\\
&-\frac{\hbar^2}{2\mu\mR^2}\times \sqrt{\mr{J}(\mr{J}+1) - \mr\Omega(\mr\Omega+1)}\times L (\mR)&
\end{eqnarray}
which are symbolized by $\mr\Gamma_{JL}^{0}$ for $\mr\Omega = 0$ and $\mr\Gamma_{JL}^{1}$ for $\mr\Omega = 1$.\\ 
Off-diagonal matrix elements of ${\mr{H}}^{\mathrm{REL}}$ are coming from ${\mr{H}}^{\mathrm{SO}}$ and ${\mr{H}}^{\mathrm{SR}}$.
From the selection rules (see \cite{Brion}), the spin-orbit interaction ${\mr{H}}^{\mathrm{SO}}$ couples the \As state to the $\mr\Omega = 0$ component of the \cs state, and leads to the strong perturbations observed in the rotational energy ladder of the A state. Also, the spin-orbit interaction couples the \as ($\mr\Omega = 1$) to the \cs ~($\mr\Omega = 1$). The contribution are: 
\begin{eqnarray}
<c \mr\Piu (0_u^+),\mr{J}|{\mr{H}}^{\mathrm{SO}}|A^1\Sigma_u^+,\mr{J}> & = & \chi(\mR)\\
<c \mr\Piu (1_u),\mr{J}\ |{\mr{H}}^{\mathrm{SO}}|\ a\tsigu(1_u),\mr{J} > & = & -\zeta(\mR)
\end{eqnarray}  
We have noted the unknown molecular matrix elements by $\chi(\mR)$ and $\zeta(\mR)$. 
Finally the non-diagonal contribution from the spin-rotation Hamiltonian is:
 \begin{equation} {\mr{H}}^{\mathrm{SR}}=(\gamma/2)(\Jop_+\Sop_- + \Jop_-\Sop_+) \end{equation}
 \begin{equation}
 \begin{split}
<\mr{J}&,\mr\Lambda,\mr{S},\mr\Sigma,\mr\Omega|{\mr{H}}^{\mathrm{SR}}|\mr{J},\mr\Lambda,\mr{S},\mr\Sigma- 1,\mr\Omega+ 1>=\\
 &\gamma(R)\sqrt{\mr{J}(\mr{J}+1) - \mr\Omega(\mr\Omega+1)} \sqrt{\mbox{S}(\mr{S}+1) - \mr\Sigma(\mr\Sigma-1)}
 \end{split}
 \end{equation}
These contributions are symbolized by $\mr\Gamma_{SR}^{0}$ for $\mr\Omega = 0$ and by $\mr\Gamma_{SR}^{1}$ for $\mr\Omega = 1$.\\ 
\twocolumn[
\tiny{ 
\begin{equation}\label{mat}
\mr H(\mR)=
\bordermatrix{ &\mr\Omega_A=0&\mr\Omega_c=0 & \mr\Omega_c=1 & \mr\Omega_a=1 & \mr\Omega_c=2\cr
 & \mr H_A^{\mr{BO}}+\JJ & \chi(\mR) & 0 & 0 & 0\cr
 &\chi(\mR)&\mr H_c^{\mr{BO}} -\gamma +\epsilon +\JJO-A(\mR)&\Gamma_{JS}^{0}+\Gamma_{SR}^{0}&\Gamma_{JL}^{0}& 0\cr
 &0&\Gamma_{JS}^{0}+\Gamma_{SR}^{0}&\mr H_c^{\mr{BO}}-2(\gamma +\epsilon)+\JJO&\Gamma_{LS}-\zeta(\mR)& \Gamma_{JS}^{1}+\Gamma_{SR}^{1} \cr
 &0&\Gamma_{JL}^{0}&\Gamma_{LS}-\zeta(\mR)&\mr H_a^{\mr{BO}} -\gamma +\epsilon +\JJ&\Gamma_{JL}^{1} \cr
 &0&0&\Gamma_{JS}^{1}+\Gamma_{SR}^{1}&\Gamma_{JL}^{1}&\mr H_c^{\mr{BO}}+A(\mR)-\gamma +\epsilon +\frac{\hbar^2}{2\mu \mR^2}(X-3)}  
\end{equation}
}
]
In summary, the contributions of the various interactions are abbreviated by:
\begin{equation}
\Gamma_{LS} = \frac{1}{2\mu\mR^2}  \sqrt{2} L(\mR)  \nonumber \\
\end{equation} 
\begin{equation}
\Gamma_{JL}^{0} = -\frac{1}{2\mu\mR^2}  \sqrt{X} L(\mR)  \ \ \ \Gamma_{JL}^{1} = -\frac{1}{2\mu\mR^2} \sqrt{X-2} L(\mR) \nonumber \\
\end{equation} 
\begin{equation}
\Gamma_{JS}^{0} = -\frac{1}{2\mu\mR^2}  \sqrt{2X}  \ \ \ \Gamma_{JS}^{1} = -\frac{1}{2\mu\mR^2} \sqrt{2X-4} \nonumber \\
\end{equation} 
\begin{equation}
\Gamma_{SR}^{0} = \frac{\gamma}{2} \sqrt{2X} \ \ \ \Gamma_{SR}^{1} = \frac{\gamma}{2} \sqrt{2X-4}  \nonumber 
\end{equation} 
We have noted $X =$ J(J+1).\\
The application to the wavefunctions given in eq. \ref{wfunc} results in the 5$\times$5 Hamiltonian matrix (\ref{mat}) for the considered subspace of states, for a given J and ($e$) symmetry. 
The matrix is ordered by the $\mr\Omega$-value of the different states $\mr\Omega_s$, where $s$ stands for the states A, c and a. H$^{\mr {BO}}_s = V^{\mr {BO}}_s(\mR) + {\mr{T}}^{\mr{N}}(\mR)$ contains the Born-Oppenheimer potentials and the kinetic energy of the relative motion. The A state potential is represented by $V_A^{\mathrm{BO}}$. 
The $\mr\Omega$-components of the c state split as $V^{\mr\Omega}_c(\mR) = V^{\mathrm{BO}}_c(\mR)$ + $(\mr\Omega -1)A(\mR)$. The potential of $\mr\Omega$ = 1  component of the a state is described by $V_a^{\mr{BO}}$. \\

\section{Global deperturbation} \label{GlobDep}

\subsection{Fourier Grid representation of the Hamiltonian} \label{FGH}

The potentials of the considered sub-space will be determined by 
minimizing the standard deviation between observed term energies and calculated term energies from solving the Schr\"odinger equation with the Hamiltonian of (eq. (\ref{mat})). The minimization is realized with a non-linear fitting routine \cite{MINUIT}.
The Schr\"odinger equation is solved using the Fourier Grid Hamiltonian method (FGH) \cite{Lisdat,Kosloff,DulieuJul}. In this method the internuclear distance is discretized by a grid with N equidistant points for a length L. In this representation the potential and coupling operators, which are local in the R-coordinate space, are diagonal. In contrast the kinetic energy matrix, which contains the derivative with respect to R is non-diagonal. Therefore the Hamiltonian matrix of a system of p coupled states has a dimension (p$\cdot$N)$\times$(p$\cdot$N), composed of p$^2$ blocks with size N$\times$N.\\
The spacing between grid points should be at least smaller than half of the smallest local de Broglie wavelength $\mr\Lambda(\mR)$ of the relative nuclear motion, following the Nyquist theorem \cite{Nyquist} for a proper calculation of the energy positions of the levels: $\mr\Delta \mR \leq \mr\pi\,\hbar/\sqrt{2\mu\mr\Delta \mr V}$, where $\mr\Delta \mr V$ is the difference between the highest considered energy and the lowest minimum of the different potentials. By calling upon the Nyquist theorem, we implicitly consider the wave function as sine waves. This is an approximation since the variation of the amplitude of the wave functions in the classically forbidden region is not sinusoidal but exponential. Therefore we multiply by a parameter lower than one to set $\mr\Delta \mR$ to lower values than estimated by the above equation for getting the uncertainty of the calculated term energies smaller than our experimental uncertainty. We found that the necessary value for a proper representation should be smaller than 0.7 times the Nyquist estimate. We used the factor 0.5 to have a correct representation and to avoid an unnecessary high number of grid points. The term energies are located between 14600 \wn and 16272 \wnf (the zero of energy is the minimum of the ground state potential see \cite{Allard2}). The classical vibrational motions are restricted in the interval from 3.08 \AA\ to 4.52 \AA\ for the A state and from 3.37 \AA\ to 8.65 \AA\ for the c state in the non-coupled picture. For our grid choice ranging from 2.12 \AA\ to 11.11 \AA\ the representation of the states of the model (p = 5) requires a $1235 \times 1235$ matrix.

\subsection{Construction of potential and coupling functions}

We split the representation of the potentials in three regions: the repulsive wall (R$<$R$_{inn}$), the asymptotic region (R$>$R$_{out}$), and the intermediate region in between.  
For the intermediate region, we use the following representation:
\begin{equation}\label{ana}
\mbox{V}_{\mathrm {fit}}(\mR)=\sum_{i=0}^{n}a_i\,\xi(\mR)^i
\end{equation}
with the analytic function
\begin{equation}\label{ana2}
\xi(\mR)=\left(\frac{\mR - R_m}{\mR + b\,R_m}\right) 
\end{equation}
where \{a$_i$\}, $b$, and $R_m$ are free parameters ($R_m$ is close to the value of the equilibrium distance). We used 13 parameters $a_i$ for the A state and 17 for the c state. The experience in our group, obtained by fitting potentials with this choice of representation, shows that this manifold of parameters should allow correct descriptions of the potential curves, for the range of observed vibrational levels and for the expected well behaved potentials according to the \textit{ab initio} calculations from \cite{Czuchaj} (no shelf or double well structure\footnote{In \cite{Czuchaj} the $^1\mr\Sigma^+_u$($^1$D +$^1$S)  is lying 2000~\wn above the $^3$P + $^1$S asymptote. The potential curve of the $^1\mr\Delta_u$($^1$D +$^1$S) state has a similar depth and shape as the A state potential determined experimentally. There is probably some wrong ordering of electronic symmetry in \cite{Czuchaj}.}).
We have no direct spectroscopic observation for the \as state, as it is weakly coupled to the \As and \cs ($\mr\Omega = 0$) states. So, we determine an initial set of parameters from the published \textit{ab initio} curve of ref. \cite{Czuchaj}.  These latter parameters were kept fixed during the fit. \\
The potentials are continuously extrapolated below $\mbox{R}_{inn}$ with:
\begin{equation}\label{rep}
\mbox{V}_{\mathrm {BO}}(\mR)= A + B/\mR^{10}
\end{equation}
by adjusting the $A$ and $B$ parameters. Beyond $\mbox{R}_{out}$ the following forms are considered respectively for the A, c and a states: 
\begin{equation}\label{LR}
\mbox{V}_{\mathrm {BO}}^{A}(\mR)= D_e^{A} - C_5^{A}/\mR^5 - C_6^{A}/\mR^6
\end{equation}
\begin{equation}\label{LR2}
\mbox{V}_{\mathrm {BO}}^{c}(\mbox{R})= D_e^{c}  - C_6^{c}/\mbox{R}^6  - C_8^{c}/\mbox{R}^8
\end{equation}
\begin{equation}\label{LR3}
\mbox{V}_{\mathrm {BO}}^{a}(\mbox{R})= D_e^{a}  - C_6^{a}/\mbox{R}^6 - C_8^{a}/\mbox{R}^8
\end{equation}
$\mR_{out}$ is different for the three states. The short and long-range extrapolation functions are connected near the shortest and the largest classical turning points of the observed levels.
The dissociation limits of the states $D_e^{A}$, $D_e^{c}$ and $D_e^{a}$ are calculated with respect to the minimum of the ground state potential \cite{Allard2},  using its value of the dissociation energy $D_e^{X} = 1102.074(9)$ \wnf:  
\begin{eqnarray}
 D_e^{A}&=& \mr\Delta E(^1D - ^1S) + D_e^{X} \nonumber\\ 
&=& 21849.634 + 1102.074=22951.708(9)\ \mbox{\wn and} \nonumber\\
D_e^{c}&=& \mr\Delta E(^3P - ^1S) + D_e^{X}  \nonumber\\
&=& 15263.003 + 1102.074 =16365.077(9)\ \mbox{\wn} \nonumber
\end{eqnarray}
The atomic transition energies are taken from reference \cite{grotrian}.\\
The parameter $C^A_5$ was fitted because the A-c coupling in the classical forbidden redion of the \As state with the classically allowed redion of the c gives significant shifts of levels. However, our value of $C^A_5$ should not be considered as true long-range coefficent because even the highest observed values are far below the atomic asymptote. The continuity through the connection point ($\mR^A_{out} = 4.511$ \AA) is ensured by $C^A_6$. The $C^c_6$ and $C^a_6$ coefficients for the \cs and the \as states were fixed to the most recent \textit{ab initio} values from \cite{PAJul}. The coefficients $C^c_8$ and  $C^a_8$ were set such that the potentials are continuous at the connecting points $\mR^c_{out}$ and $\mR^a_{out}$.\\  

\subsection{Four-states model} \label{4s}

\begin{figure}
\centering
\resizebox{0.95\columnwidth}{!}{
  \includegraphics{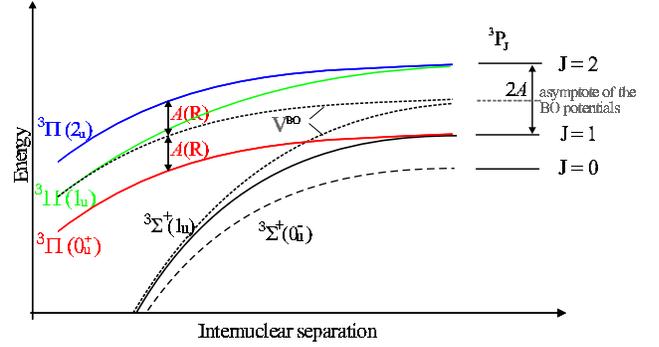}}
\caption{Adiabatic potentials of u-symmetry states correlated to the $^3$P + $^1$S asymptote. \label{asymp} } 
\end{figure}

Our data set contains only levels, which are accessible via the ground state, i.e. which have a strong singlet character ($> 15 \%$ an estimation from the deperturbation analysis below). Since the $\mr\Omega=1$ component of the \cs and \as states and the $\mr\Omega = 2$ component of the \cs state are weakly coupled to the \As state, the number of levels of these states which have a significant singlet character is low. Hence, the information about the $\mr\Omega=1$ and $2$ components of the \cs state is low or negligible with our data set. Therefore, the information to characterize the spin-orbit splitting $A$(R) is limited. We then define the fitted potential as the potential of the $\mr\Omega = 0$ component of the \cs state:
\begin{equation}
V^{\mr{fit}}_c(\mR) = V^{0}_c(\mR)
\end{equation}
Consequently the potentials for the other $\mr\Omega$-components of the c state are defined as:
 \begin{equation}
V^{\mr{\Omega}}_{mod}(\mR) = V^{\mr{fit}}_c(\mR) + \mr\Omega A(\mR)
\end{equation}
The \as state potential is about four times deeper than the investigated energy range below the asymptote $^3$P+$^1$S, which induces a small grid step in the FGH representation (see section \ref{FGH}), and increases the Hamiltonian matrix size in the FGH representation. Since this state is weakly coupled to the \cs ($\mr\Omega = 0$) and only strongly coupled to the \cs ($\mr\Omega = 1$) its influence on the description of the spectroscopic data is expected to be weak. We took into account the coupling between the \as ($\mr\Omega =$ 1) state and the \cs ($\mr\Omega =$ 1) state by replacing the $\mr c^3\mr\Pi(\mr\Omega=1)$ potential by the adiabatic $1_u(\mr c^3\mr\Pi)$ potential with respect to the spin-orbit coupling $\zeta$(R) (fig. \ref{asymp}), 
obtained by the diagonalization of the simplified 2$\times$2 sub-matrix from eq. (\ref{mat}) for each R:
\begin{equation} \label{mat2}
\centering
\bordermatrix{&\mathrm{\mr\Omega}_c = 1&\mr\Omega_a = 1 \cr
&V_c^{\mathrm{fit}} + A(\mR) &\mr\Gamma_{LS}-\zeta(\mR)\cr
&\mr\Gamma_{LS}-\zeta(\mR) &V_a^{BO}\cr}
\end{equation}
where rotational couplings have been neglected. The model is thus reduced to an effective 4-states model, and consequently the total representation is reduced to a $293\times 293$ matrix. The spin-orbit coupling $\zeta(\mR)$ has been approximated by $\zeta(\mR)=A(\mR)$.  \\
To our knowledge, there are no \textit{ab initio} calculations available, which could properly model the spin-orbit splitting $A$(R) and the spin-orbit couplings $\chi$(R). Only the values at the atomic limit are known. The spin-orbit splitting converges to the atomic spin-orbit value $A_0 = 52.940$~\wn \cite{grotrian}. The spin-orbit coupling $\chi$(R) vanishes for large R since the atomic coupling between the (4s3d)$^1$D state and the (4s4p)$^3$P state is zero because of different parities. We have represented these interactions in different R ranges as for the potentials (see above). Polynomial functions in the inner range allow for sufficient flexibility. For the outer region we ensure continuity at the connecting point R$_c$ and correct atomic values for the asymptote. The representation of the functions is expressed as:
\begin{equation}
\label{Coupform}
\kappa(\mR) = \left\{
\begin{array}{l@{\;\;}l@{\;\;}l}
\kappa(R_\mr{s}) + \kappa_\mr{s} \times (\mR-\mR_\mr{s})^2 & \mbox{\rm for} & \mR < \mR_s \\
\sum_{p=0}^{N} \kappa_p(\mR-\mR_2)^p                       & \mbox{\rm for} & \mR_s \le \mR < \mR_c \\
\kappa^o_1/\mR^1 + \kappa^o                                & \mbox{\rm for} & \mR_c \le \mR \\
\end{array}
\right.
\end{equation}
where $\kappa(R)$ stands for $A$(R) or $\chi$(R), and asymptotically $\kappa^o = A_0$ for $A$(R), and $\kappa^o = 0$ for $\chi$(R). The expansion parameter R$_2$ is chosen to be close to the internuclear distance corresponding to the crossing between the \As state and the \cs state, and $\kappa_p$, $\kappa_s$ are fitting parameters. The parabolic extrapolation below R$_{\mr s}$ is used only for $\chi$(R). $A$(R) is not truncated below R$_{\mr s}$. 

\subsection{Fitting strategy}

\begin{figure}
\centering
\resizebox{0.95\columnwidth}{!}{
  \includegraphics{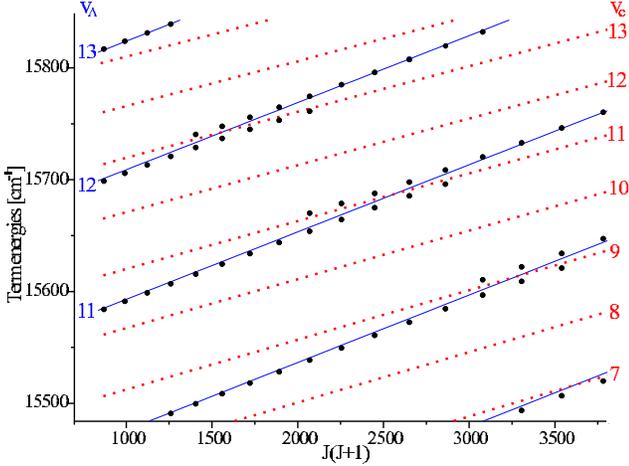}}
\caption{Observed (full circle) and calculated initial term energies for the c state (dotted lines) and for the A state (full lines) as function of J(J+1) for the same region as the insert figure \ref{term}. Vibrational quantum numbers are given at the left for the A state and at the right for the c state.}  
\label{gessl}
\end{figure}
The fit of our spectroscopic data to analytical potential curves and coupling functions is a highly non-linear problem and the parameters to be derived are strongly correlated. Therefore, we should start the fitting procedure with the simplest model possible in order to reduce the number of parameters and thus the complexity of the problem. For this reason, we first limited the model to the $\mr\Omega = 0^+$ components of the \As state and the \cs state and simplified the polynomial part of the non-diagonal spin-orbit functions (eq. (\ref{Coupform})) to an adjustable constant.   \\
Additionally, the time for achieving convergence of the fit depends critically on the initial guesses for the potential energy curves and the couplings. Therefore, we constructed initial potentials for the $\mr\Omega = 0^+$ components of the \As and the \cs states, which yield the best possible reproduction of the observed rovibrational ladders. As we can see in the figures \ref{term} and \ref{gessl}, the vibrational and rotational structure can be well identified for the A state. To construct the potentials, we calculated traditional Dunham parameters from the rovibrational ladder and derived RKR potentials with these parameters and converted to analytic potentials according to eq. (\ref{ana}) by a simple linear fit. For the c state, we only have the local perturbations to construct an initial potential. The rotational ladder of one vibrational level of the c state crosses several rotational ladders of different vibrational levels of the A state at different J-values. This was used to estimate the rotational constants, and vibrational spacing to obtain an initial c state potential. These constructed potentials give the rovibrational levels exemplified in figure \ref{gessl}. In this manner, we ensured that the ladders of both states cross at the observed local perturbations. \\ 
We can expect that the coupling of the states leads to a shift of each level locally caused by the closest levels of the perturbing state and to a global shift coming from the accumulated influence of all the other levels. The selection of levels following a regular rovibrational series estimated from the observed levels, which are not strongly deviating from such regular behavior, gives a rough estimation of the positions of the levels of the uncoupled states, since the global effect is not properly taken into account. 
In a first iteration step, we fitted only the spin-orbit constant.  In this way, we obtained a first guess of the value of this constant and the magnitude of the shifts caused by the coupling. These shifts have been subtracted from the observed term energies, which are not strongly perturbed. Then, the initial potential of the A state has been fitted to these roughly deperturbed levels. The obtained potential has been used again to improve the value of the spin-orbit constant. The procedure has been repeated several times to improve iteratively the A state potential and the coupling. Then, this method has been employed once for improving the c state using this time the strongly perturbed levels and the obtained value of the coupling constant. 
The initial potentials and the constant coupling result in a dimensionless standard deviation $\bar\sigma = 170$ between observed and calculated levels. The improved potentials obtained by the iterative procedure lead to a standard deviation of $\bar\sigma = 60$. \\
The obtained potentials and the value of the spin-orbit parameter provided the starting conditions for the global fit itself.

\subsection{Result with the 4-states model}

Practically, we have extended the fit iteratively, first both $0^+$ components, second adding $\mr\Omega = 1$ of c$^3\mr\Pi$ and third $\mr\Omega = 2$ of c$^3\mr\Pi$ to avoid to fit directly a large number of free parameters. This is justified since the magnitude of the coupling is weak between components with $\mr\Omega \neq 0$ and $\mr\Omega = 0$ compared to the spin-orbit coupling between the \As state and the \cs ($\mr\Omega =0$) state. Therefore, the adiabatized $\mr\Omega =1$ component and the $\mr\Omega =2$ component of the c state and the diagonal spin-orbit splitting were successively introduced when the convergence of the simplified models was reached. The only contribution, which remains neglected in the model, is the spin-spin splitting (eq. (\ref{SS})) since its magnitude is expected to be small compared to our experimental uncertainty. The parameter $\epsilon$ was kept at zero.\\
\begin{figure}
\centering
\resizebox{1.0\columnwidth}{!}{
  \includegraphics{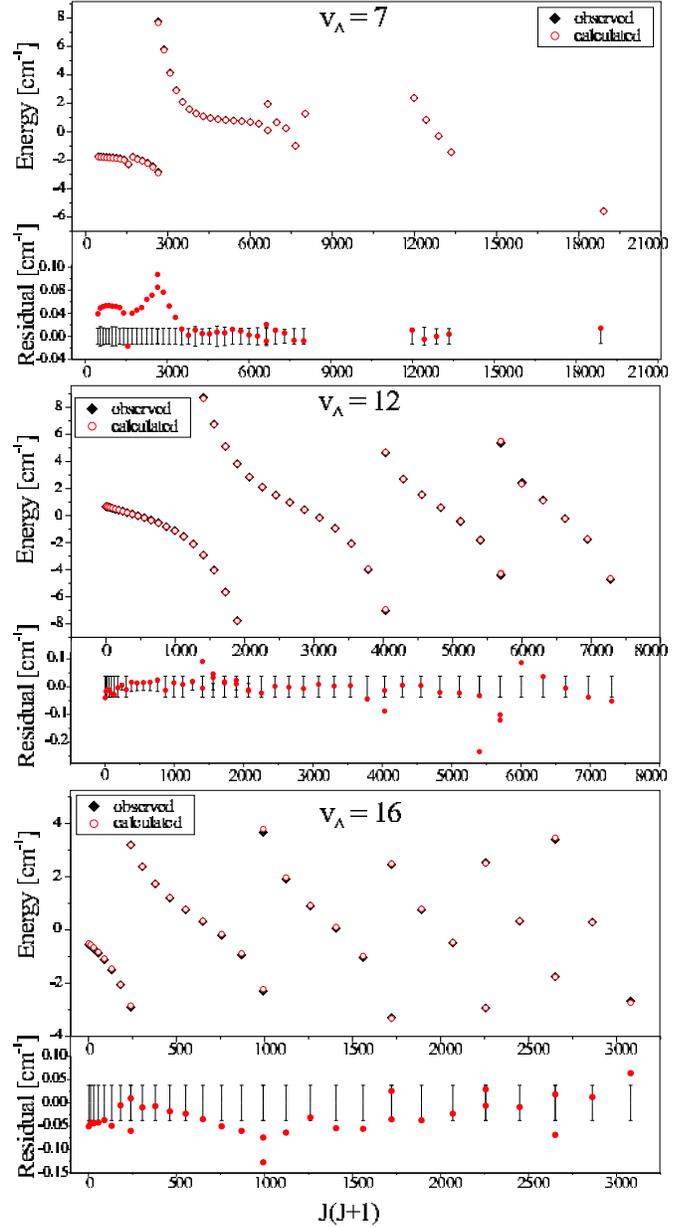}}
\caption{The calculated term energies with the 4-states model and observed term energies as function of J(J+1) for three selected vibrational levels \vpA = 7, 12, 16 are presented. A linear rotational function E$_v$(J = 0) + B$_v$$\times$[J(J+1)] has been subtracted from the term energies in order to show the deviations from the regular rotational series due to the perturbations. The lower graphs present the difference between observed and calculated term energies and the experimental error bars.} 
\label{result}
\end{figure}
Figure \ref{result} shows the quality of the final representation of the observed levels, which have been reached with the 4-states model. The rotational ladders of \vpA = 7, 12, and 16 are taken as examples. For each case in the upper part, we show the observed levels from which we have subtracted the linear function E$_v$(J = 0) + B$_v$ $\times$[J(J+1)] in order to show better the good reproduction of the perturbations. For each \vpA, the lower graph shows the differences between observed and calculated term energies. We see that most of the residuals lie within the experimental error bars. The perturbations, which can reach magnitudes of 8 \wnf, are reproduced to tenths of a wave number. Therefore, the global reproduction of the position of observed levels given by the normalized standard deviation $\sigma = 1.65$ is satisfactory corresponding to a standard deviation of 0.053 \wnf. This quality is only achievable with a model which includes the four considered states. Figure \ref{residuals7} shows the improvement beginning with the 2-states model, containing only the components $\mr\Omega = 0^+$ of the \As and \cs states, by the successive addition of the adiabatized $\rm\Omega = 1$ component (3-states model) and $\rm\Omega = 2$ component (4-states model) of the \cs state. 
The improvement is particularly visible for the A state levels near J = 39 and 81, which have a clear mixing with the $\rm\Omega = 1$ state and a weaker mixing with $\rm\Omega = 2$. Note the different scales of the panels. The inclusion of the third and, later on, fourth state leads to local improvements but also gives a better overall representation. The relative magnitudes of the residuals for the 2-states model compared to the 3-states model exemplify clearly the important role of the different states. The achieved standard deviation with the 2-states model is only $\sigma = 4$ and $\sigma = 1.98$ for the 3-states model, and finally $\sigma = 1.65$ for the 4-states model.\\
Nevertheless, a completely satisfying reproduction of the observed levels within their experimental uncertainties was not reached with the considered model. 
\begin{figure}
\centering
\resizebox{1.0\columnwidth}{!}{
  \includegraphics{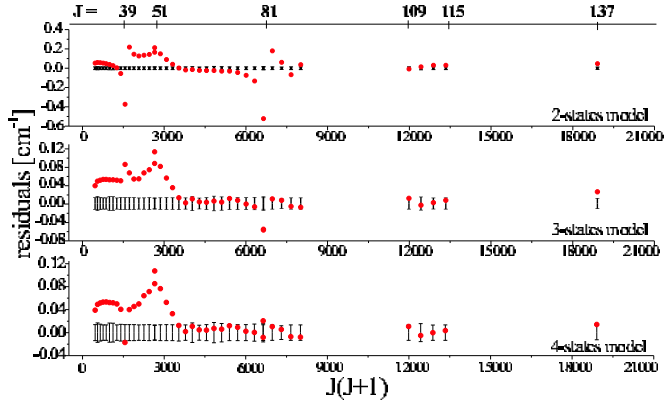}}
\caption{Residuals as function of J(J+1) obtained for the rotational series of \vpA = 7 with the 2, 3 and 4 - states models. The corresponding J quantum numbers at the local perturbations are given on top of this figure.} 
\label{residuals7}
\end{figure}
In figures \ref{result} and \ref{residuals7}, we see that systematic trends in the residuals remain (for instance in fig. \ref{result} for \vpA = 16 between J(J+1)$\approx 250$ and $1000$ ). We also see that the large residuals are located systematically at the position of the strong perturbations and they are two to four times larger than the error bars. We verified that such trends and large residuals will not be reduced by an increase of the number of parameters for the potentials. \\ 
\begin{figure}
\centering
\resizebox{0.95\columnwidth}{!}{
  \includegraphics{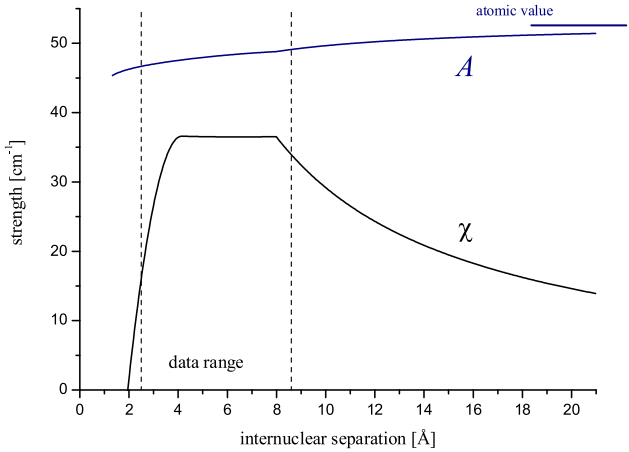}}
\caption{Fitted spin-orbit splitting $A$(R) of c$^3\mr\Pi$ and spin-orbit coupling $\chi$(R) of \As and \cs($0^+$).} 
\label{coup}
\end{figure}
Figure \ref{coup} shows the spin-orbit functions obtained by the fit. In the range of data (i.e. within the classical turning points of observed levels), their variations are slow, despite the large flexibility that the choice of representation (eq. (\ref{Coupform})) offers. Therefore, an increase of the number of parameters for the coupling functions will not give a better representation of the observations. This can be explained by the fact that the data are mainly sensitive to the averaged strength of the couplings and not so much to the details of the shape of the couplings. Hence, obtaining more accurate spin-orbit functions from \textit{ab initio} calculations may not help to improve the fit using the present data set.\\
The derived parameters for the potentials and the coupling functions are given in tables \ref{tabA} to \ref{tabtcc}. The potential of the \As state is determined in the internuclear interval from 3.08 \AA\ to 4.52 \AA\ and the fitted potential for the \cs from 3.37~\AA\ to 8.60~\AA. The outer turning points for the two potentials are located at relatively short distances corresponding to potential points located 6713 \wn below the $^1$D + $^1$S asymptote for the \As state and 40 \wn below the $^3$P + $^1$S asymptote for the \cs(0$^+$) state. Consequently, the dispersion coefficients given in table \ref{tabA} to \ref{tabta} should be considered only as extrapolating parameters in order to have proper boundary conditions to diagonalize the Hamiltonian matrix.   

\begin{table}[ht]
\fontsize{8pt}{13pt}\selectfont
\caption{Parameters of the analytic representation of the \As state potential. The energy reference is the minimum of the ground state potential. Parameters with $^\ast$ are set for extrapolation of the potential.\label{tabA}}
\begin{tabular*}{1.0\columnwidth}{@{\extracolsep{\fill}}|lr|} 
\hline
   \multicolumn{2}{|c|}{$R \leq R_\mathrm{inn}=$ 3.080 \AA}    \\
\hline
   $A^\ast$ & 0.131237254$\times 10^{5}$ \wn \\
   $B^\ast$ & 0.23850904$\times 10^{9}$  \wn \AA $^{10}$ \\
\hline
   \multicolumn{2}{|c|}{3.080 \AA\ $< R < R_\mathrm{out}=$ 4.511 \AA}    \\
\hline
    $b$ &   $-0.57$              \\   
    $R_\mathrm{m}$ & 3.59402020 \AA               \\
    $a_{0}$ &  14106.8528 \wn\\
    $a_{1}$ & -0.2251508202149158$\times 10^{2}$ \wn\\
    $a_{2}$ &  0.1220394308918707$\times 10^{5}$ \wn\\
    $a_{3}$ &  0.6764477900769186$\times 10^{4}$ \wn\\
    $a_{4}$ & -0.4630366850561168$\times 10^{4}$ \wn\\
    $a_{5}$ &  0.1699400434605241$\times 10^{5}$ \wn\\
    $a_{6}$ &  0.9375762656463659$\times 10^{5}$ \wn\\
    $a_{7}$ & -0.4561172682499647$\times 10^{5}$ \wn\\
    $a_{8}$ & -0.6354314702263585$\times 10^{6}$ \wn\\
    $a_{9}$ & -0.6783215663449778$\times 10^{6}$ \wn\\
   $a_{10}$ &  0.1208346212864403$\times 10^{7}$ \wn\\
   $a_{11}$ &  0.3797632603225922$\times 10^{7}$ \wn\\
   $a_{12}$ &  0.3059003584697336$\times 10^{7}$ \wn\\
\hline
   \multicolumn{2}{|c|}{$R_\mathrm{out} \leq R$}\\
\hline              
  ${D^{\mathrm A}_{\mathrm e}}$ & 22951.708(9) \wn    \\
  ${C_5}$ &  0.5028185$\times 10^{8}$ \wn\AA$^5$      \\
 ${C_{6}}^\ast$ & -0.1699751$\times 10^{9}$ \wn\AA$^6$   \\
\hline
 \multicolumn{2}{|c|}{Additional constants:} \\
\hline
\multicolumn{2}{|l|}{equilibrium distance:\hspace{2.2cm} $R_e^A$= 3.595(1) \AA} \\
\multicolumn{2}{|l|}{electronic term energy:\hspace{1.6cm}     $T_e^A$= 14106.8(10) \wn}\\
\hline
\end{tabular*}
\end{table}

\begin{table} 
\fontsize{8pt}{13pt}\selectfont
\caption{Parameters of the analytic representation of the \cs(0$^+$) state potential. The energy reference is the minimum of the ground state potential. Parameters with $^\ast$ are set for extrapolation of the potential. \label{tabc}}
\begin{tabular*}{1.0\columnwidth}{@{\extracolsep{\fill}}|lr|}
\hline
   \multicolumn{2}{|c|}{$R \leq R_\mathrm{inn}=$ 3.319 \AA}    \\
\hline
   $A^\ast$ & 0.146530549$\times 10^{5}$ \wn \\
   $B^\ast$ & 0.29313693$\times 10^{9}$  \wn \AA $^{10}$ \\
\hline
   \multicolumn{2}{|c|}{3.319 \AA\ $< R < R_\mathrm{out}=$ 8.450 \AA}    \\
\hline
    $b$ &   $-0.22$              \\   
    $R_\mathrm{m}$ & 4.06654500 \AA               \\
    $a_{0}$ &  14841.9738 \wn\\
    $a_{1}$ & -0.3102692708740178$\times 10^{1}$ \wn\\
    $a_{2}$ &  0.1682036434942518$\times 10^{5}$ \wn\\
    $a_{3}$ & -0.1609245688758938$\times 10^{5}$ \wn\\
    $a_{4}$ &  0.3095300726978432$\times 10^{5}$ \wn\\
    $a_{5}$ & -0.4053968648443735$\times 10^{6}$ \wn\\
    $a_{6}$ & -0.2549434419469327$\times 10^{7}$ \wn\\
    $a_{7}$ &  0.2063831175657142$\times 10^{8}$ \wn\\
    $a_{8}$ &  0.1659840951559789$\times 10^{8}$ \wn\\
    $a_{9}$ & -0.3493050617277784$\times 10^{9}$ \wn\\
   $a_{10}$ &  0.3805220943990989$\times 10^{9}$ \wn\\
   $a_{11}$ &  0.2346473224525490$\times 10^{10}$ \wn\\
   $a_{12}$ & -0.5935661971035855$\times 10^{10}$ \wn\\
   $a_{13}$ & -0.1908247395049612$\times 10^{10}$ \wn\\
   $a_{14}$ &  0.2221570121722804$\times 10^{11}$ \wn\\
   $a_{15}$ & -0.2838711585484613$\times 10^{11}$ \wn\\
   $a_{16}$ &  0.1186121797412387$\times 10^{11}$ \wn\\
\hline
   \multicolumn{2}{|c|}{$R_\mathrm{out} \leq R$}\\
\hline              
 ${D^{\mathrm c}_{\mathrm e}}$(0$^+$) & 16312.139(10) \wn    \\
 ${C_{6}}$ from \cite{PAJul} &  0.1186500$\times 10^{8}$ \wn \AA $^6$   \\
 ${C_{8}}^\ast$ & 0.3427436$\times 10^{9}$ \wn\AA$^8$   \\
\hline
 \multicolumn{2}{|c|}{Additional constants:} \\
\hline
\multicolumn{2}{|l|}{equilibrium distance:\hspace{2.2cm} $R_e^c$=4.067(1) \AA} \\
\multicolumn{2}{|l|}{electronic term energy:\hspace{1.6cm}     $T_e^c$= 14842.0(10) \wn}\\
\hline
\end{tabular*}
\end{table}

\begin{table} 
\fontsize{8pt}{13pt}\selectfont
\caption{Parameters of the analytic representation of the \as state potential from the \textit{ab initio} potential published in \cite{Czuchaj}. The energy reference is the minimum of the ground state potential. Parameters with $^\ast$ are set for extrapolation of the potential.\label{tabta}}
\begin{tabular*}{1.0\columnwidth}{@{\extracolsep{\fill}}|lr|} 
\hline
   \multicolumn{2}{|c|}{$R \leq R_\mathrm{inn}=$ 2.900 \AA}    \\
\hline
   $A^\ast$ & 0.100173775$\times 10^{5}$ \wn \\
   $B^\ast$ & 0.25535671$\times 10^{9}$  \wn \AA $^{10}$ \\
\hline
   \multicolumn{2}{|c|}{2.900\AA\ $< R < R_\mathrm{out}=$ 8.000 \AA}    \\
\hline
    $b$ &   $-0.21$              \\   
    $R_\mathrm{m}$ & 3.74686320 \AA               \\
    $a_{0}$ &   8476.274656 \wn\\
    $a_{1}$ & -0.4989163630417485$\times 10^{2}$ \wn\\
    $a_{2}$ &  0.5159297957888653$\times 10^{5}$ \wn\\
    $a_{3}$ & -0.8701696364839145$\times 10^{4}$ \wn\\
    $a_{4}$ & -0.5354588124673624$\times 10^{5}$ \wn\\
    $a_{5}$ &  0.9811459564252647$\times 10^{4}$ \wn\\
    $a_{6}$ & -0.9842985554356317$\times 10^{5}$ \wn\\
    $a_{7}$ & -0.6320055597858555$\times 10^{6}$ \wn\\
    $a_{8}$ &  0.7170441797704500$\times 10^{6}$ \wn\\
    $a_{9}$ &  0.2295720046533766$\times 10^{7}$ \wn\\
   $a_{10}$ & -0.2613404819600489$\times 10^{7}$ \wn\\
\hline
   \multicolumn{2}{|c|}{$R_\mathrm{out} \leq R$}\\
\hline              
  ${D^{\mathrm a}_{\mathrm e}}$ & 16365.078(9) \wn    \\
 ${C_{6}}$ from \cite{PAJul} &  0.1855883$\times 10^{8}$ \wn\AA $^6$   \\
 ${C_{8}}^\ast$ &  0.4785476$\times 10^{9}$ \wn \AA $^8$   \\
\hline
 \multicolumn{2}{|c|}{Additional constants:} \\
\hline
\multicolumn{2}{|l|}{equilibrium distance:\hspace{2.2cm} $R_e^a$=3.748 \AA} \\
\multicolumn{2}{|l|}{electronic term energy:\hspace{1.6cm}     $T_e^a$= 8476.3 \wn}\\
\hline
\end{tabular*}
\end{table}

\begin{table}
\centering 
\fontsize{8pt}{13pt}\selectfont
\caption{Parameters of the analytic representation of the coupling functions $A$(R), $\zeta$(R) and $\chi$(R) as defined eq. \ref{Coupform}, eq. \ref{SS1} and \ref{gamma0}. \label{tabtcc}}
\begin{tabular}{@{\extracolsep{\fill}}|c|l|} 
\hline
 coupling function &\multicolumn{1}{|c|}{parameter}    \\
\hline\hline

\multirow{11}{1.7cm}{$\chi$(R)} 
 & \multicolumn{1}{|c|}{R $<$ R$_\mr{s}$ = 4.149 \AA }   \\  
\cline{2-2}
 & $\chi_{\mr s}$ = $-0.756636\times 10^{1}$ \wn\AA $^{-2}$\\
\cline{2-2} 
 & \multicolumn{1}{|c|}{R$_\mr{s}$ $\leq$ R $<$ R$_\mr{c}$ = 8.000 \AA  }   \\  
\cline{2-2}
 & R$_2$ = 6.00 \AA \\
 & $\chi_0$ = $0.364981\times 10^{2}$ \wn \\
 & $\chi_1$ = $-0.177038\times 10^{-1}$ \wn\AA $^{-2}$ \\
 & $\chi_2$ = $0.186890\times 10^{-1}$ \wn\AA $^{-2}$ \\
\cline{2-2}
 & \multicolumn{1}{|c|}{ R$_\mr{c}$ $\le$ R }   \\ 
\cline{2-2} 

 &  $\chi^o_1$ = $0.104382 \times 10^{4}$ \wn\AA \\
 &  $\chi^o$ = $0$ \wn \\        
\hline
\multirow{8}{1.7cm}{$A$(R) = $\zeta$(R)} 
 & \multicolumn{1}{|c|}{R $<$ R$_\mr{c}$ }   \\  
\cline{2-2}
 & R$_2$ = 6.00 \AA \\ 
 & A$_0$ = $0.483295\times 10^{2}$ \wn \\
 & A$_1$ = $0.294419$ \wn\AA $^{-1}$ \\
 & A$_2$ = $-0.2914325\times 10^{-1}$ \wn\AA $^{-2}$ \\
\cline{2-2}
 & \multicolumn{1}{|c|}{R$_\mr{c}$ $\le$ R}   \\ 
\cline{2-2}  
 &  A$^o_1$ = $-0.118252\times 10^{3}$ \wn\AA \\
 &  A$^o$ = $0.529400\times 10^{2}$ \wn \ \ from \cite{grotrian}\\      
\hline
$\gamma$ & $0.1632\times 10^{-2}$  \wn \\
$\epsilon$ & fixed to zero\\
\hline

\end{tabular}
\end{table}

\section{Discussion} \label{Disc}

The variations of the residuals seen fig. \ref{result} and fig. \ref{residuals7} could be attributed to perturbations caused by other states. But they have small magnitudes and appear at positions of level crossings between the \As state and the two components ($\mr\Omega =$ 0 and 1) of the \cs states. Therefore, possible additional perturbing states should only have an indirect coupling to those latter states or a direct coupling to the \cs ($\mr\Omega = 1$) state. According to the \textit{ab initio} calculations from Czuchaj \textit{et al}. \cite{Czuchaj} the bottom of the $^3\mr\Pi_u$ state and the $^3\mr\Delta_u$ state, dissociating to $^3$D + $^1$S, approach region of observed levels. That $^3\mr\Pi_u$ state is coupled directly to the A state and to all $\mr\Omega$-components of the c state. Thus, perturbations should also occur at different J and not only at J corresponding to rotational level crossings between the A state and the c state. Such additional perturbations are not observed within our experimental uncertainty. This state is probably less deep than predicted and may lead only to a global shift to all the levels. The derived potentials in our model could include such global shift and would be in this respect effective potentials.\\
The $^3\mr\Delta_u$($\mr\Omega = 1$) state and the \as state are indirectly coupled to the $\mr\Omega = 0$ components of the A and c states via a direct coupling to the c$^3\mr\Pi_u$($\mr\Omega = 1$) state by spin-orbit interaction. We took the potential curves of these states \cite{Czuchaj}, one by one, and the non-diagonal spin-orbit function $\chi$(R) to calculate their respective influence on the position of levels of the coupled states, leading to a 5-states model. We found that among these states only the $^3\mr\Delta_u$ can create additional deviations to the calculated levels. They appear at positions of strong perturbations between the levels of the \As and \cs ($\mr\Omega = 0$) and at positions of weaker perturbations between the \As states and the \cs ($\mr\Omega = 1$) state. Furthermore, the calculated magnitude of these perturbations is of the same order as the observed larger residuals. 
The $^3\mr\Delta_u$ is a good candidate to explain why the present model was incomplete for providing a totally satisfactory reproduction of the observed levels. For reaching the experimental precision by the coupled-states calculation it might be thus necessary to extend the 4-states model by the $^3\mr\Delta_u$($\mr\Omega = 1$) and ($\mr\Omega = 2$) states. This implies that spectroscopic data for these states are needed, which could be obtained by multi-photon spectroscopy or by searching for very weak extra lines from the coupling. \\
Notable differences exist with the recent \textit{ab initio} calculations of potential energy curves for these states, in particular for the \cs(0$^+$) state. The results of the relativistic \textit{ab initio} calculations presented in \cite{NIST} give a dissociation energy of 168 \wnf, which is in strong disagreement with the value of 1470.2(10) \wn determined in our experimental study. Furthermore, the last two potential points given in \cite{NIST} at large internuclear distance lie above the expected asymptotic limit, which is in contradiction with the attractive long-range behavior of the \cs(0$^+$) state (see \cite{PAJul}). The dissociation energy of 1575 \wn of the \cs state derived with non-relativistic calculations by Czuchaj \textit{et al.} \cite{Czuchaj} is in reasonable agreement (within 6.6\%) with our experimental value.  The agreement with the dissociation energy of the \As of 9518 \wn given in \cite{Czuchaj} (Remember the proposed reassignment given in footnote 1) compared to our derived experimental value of 8845.0(10) \wn is in the same order (within 7.6\%).

\section{Conclusion} \label{Conc}

We have extended the spectroscopy \cite{Bond84,Hof84,Hof86} of the coupled system \As($^1$D + $^1$S), \cs($^3$P + $^1$S) and \as($^3$P + $^1$S) using the laser induced fluorescence and the filtered laser excitation technique. Rotational series of six new vibrational levels at the bottom of the A state have been observed compared to the previous study of Hofmann and Harris \cite{Hof86}. It allowed us to reassign the A state with great confidence. The lowest perturbations of the A state levels have been identified and consequently the c state has been reassigned. We have developed a description of the observed levels in which the spin-orbit couplings and the rotational couplings between the \As, \cs and \as are modeled. The Fourier Grid Hamiltonian method included in a fitting procedure allowed a global treatment of the coupled states, and led to a detailed description of the spectroscopic observations compared to the previous derivation from Hofmann and Harris \cite{Hof86}. They have obtained a reproduction of the observed 340 term energies of 0.13 \wn by a local deperturbation analysis. 183 coupling parameters and 16 Dunham coefficients were determined to reach this quality. These large numbers are due to the fact that the coupling of each level to all the other levels of the coupled states are modeled with local perturbations introducing for each new perturbing level a new parameter. In addition, there is some ambiguity in the choice of levels to be considered as influencing. This shows the limits of local methods for the treatment of relatively strongly coupled states. \\
We have determined potential energy curves for the \As state and the \cs state as well as spin-orbit coupling functions. For the two effective states (\As, \cs($0^+$)) and the coupling functions 39 parameters have been derived from the fit and 22 parameters fixed for the \as(1) state potential or for the continuation to small or large R . 
Small deviations remain with the present model. We believe that such effects can be explained by the neighboring $^3\mr\Delta_u$($^3$D + $^1$S) state, which was not reasonable to include in the model. More precise data of the studied system, and in particular for the \cs state and the $^3\mr\Delta_u$ are needed. \\
In their spectroscopic investigations of Ca$_2$ in a supersonic jet Bondybey and English \cite{Bond84} have observed band system centered around 15000 \wn belonging to the \As $-$ \Xs system and additional bands that were not identified (see fig. 1 in \cite{Bond84}). Our analysis allowed us to correct the vibrational assignment of these molecular bands and to assigned the previously unidentified bands to the \cs $-$ \Xs system (see \cite{ThesisAllard}). Since levels of the \cs state have been directly observed in their jet experiment, the spectroscopic study of the c state in a molecular beam seems promising. Particularly the observation of high lying levels of the c state is of interest for a better characterization of the collision processes between one $^3$P and one $^1$S calcium atom. The short-range coupling of the other state, except the \as state, are predicted to be weak for the long-range levels of the c state. In this respect the descriptions will not involve a more complex model than the present one. Therefore, this model should provide a good understanding of ultra-cold collisions in a Ca trap where the intercombination line of calcium is studied to establish this resonance as an optical frequency standard. \\

\section*{Acknowledgments}

This work was supported by DFG through SFB 407 and in part by the European Union in the frame of the Cold Molecules TMR Network under contract No. HPRN-CT-2002-00290. A. P. gratefully acknowledges a research sti\-pend from the Alexander von Humboldt Foundation.


\begin{thebibliography}{9}
\bibitem{CaPTB}
G. Wilpers, T. Binnewies, C. Degenhardt, U. Sterr, J. Helmcke, and F. Riehle, Phys. Rev. Lett. \textbf{89}, (2002) 230801.
\bibitem{CaNIST}
E. A. Curtis, C. W. Oates, and L. Hollberg, J. Opt. Soc. Am. B -Opt. Phys. 20, (2003) 977.
\bibitem{Allard1} 
O. Allard, A. Pashov, H. Kn\"ockel, and E. Tiemann, Phys. Rev. A \textbf{66}, (2002) 042503.
\bibitem{Allard2}
O. Allard, C. Samuelis, A. Pashov, H. Kn\"ockel, and E. Tiemann, Eur. Phys. J. D \textbf{26}, (2003) 155.
\bibitem{Lisdat}
Ch. Lisdat, O. Dulieu, H. Kn\"ockel, and E. Tiemann, Eur. Phys. J. D \textbf{17},(2001) 319.
\bibitem{ThesisAllard}
O. Allard, \textit{Long-range interaction in the calcium dimer studied by molecular spectroscopy}, thesis, University of Hannover and Paris XI (2004).
\bibitem{Bond84}
V. E. Bondybey and J.H. English, Chem. Phys. Lett \textbf{111}, (1984) 195.
\bibitem{Hof84}
R. T. Hofmann and D. O. Harris, J. Chem. Phys. \textbf{81}, (1984) 1047.
\bibitem{Hof86}
R. T. Hofmann and D. O. Harris, J. Chem. Phys. \textbf{85}, (1986) 3749.
\bibitem{Iod}
IodineSpec program can be found on the web site www.toptica.com.
\bibitem{Brion}
H. Lefebvre-Brion and R.W. Field, \textit{Perturbation in the spectra of diatomic molecules}, Harcourt Brace Jovanovich (1986).
\bibitem{Czuchaj}
E. Czuchaj, M. Krosnicki, and H. Stoll, Theo. Chem. Accounts \textbf{110}, (2003) 28.
\bibitem{Brown}
J. M. Brown, J. T. Hougen, K.-P. Huber, J. W. C. Johns, I. Kopp, H. Lefebvre-Brion, A. J. Merker, A. J. Merer, D. A. Ramsay, J. Rostas, and R. N. Zare, J. Mol. spectrosc. \textbf{55}, (1975) 500.
\bibitem{MINUIT}
MINUIT, CERN program library long writeup D506, wwwasdoc.web.cern.ch/minuit/minmain.html.
\bibitem{Kosloff}
R. Kosloff, J. Phys. Chem. \textbf{92}, (1988) 2087.
\bibitem{DulieuJul}
O. Dulieu and P.S. Julienne, J. Chem. Phys. \textbf{103}, (1995) 60.
\bibitem{Nyquist}
W. H. Press et al., \textit{Numerical Recipes}, Cambridge University Press (1987).
\bibitem{grotrian}
S. Bashkin and J.O. Stoner, \textit{Atomic Energy-Levels and Grotrian Diagrams}, 2, (North-Holland, Amsterdam, 1978).
\bibitem{PAJul}
R. Ciurylo, E. Tiesinga, S. Kotochigova, and P.S. Julienne, Phys. Rev. A, \textbf{70}, (2004) 062710.
\bibitem{NIST}
S. Kotochigova and P.S. Julienne, Potential Energy Surface Database of Group II Dimer Molecules, physics.nist.gov/PhysRefData/PES/index.html, National Institute of Standards and Technology, USA.
\end{thebibliography}
\end{document}